\documentstyle [12pt]{article}
\input amssym.def
\input amssym.tex
\begin{document}

\begin{titlepage}

\title{General solutions of the  Monge-Amp\`{e}re equation in $n$-dimensional
space}

\author{D.B. Fairlie\\
{\it Department of Mathematical Sciences}\\
{\it University of Durham, Durham DH1 3LE}\\
and A.N. Leznov\\
{\it  Institute for High Energy Physics, 142284 Protvino,}\\{\it Moscow Region,
Russia}}
\maketitle

\begin{abstract}

It is shown that the general solution of a homogeneous Monge-Amp\`{e}re
equation in
$n$-dimensional space is closely connected with the exactly (but only
implicitly)
integrable system
\begin{equation}
\frac {\partial \xi_{j}}{\partial x_0}+\sum_{k=1}^{n-1}
\xi_{k} \frac {\partial \xi_{j}}{\partial x_{k}}=0 \label{1}
\end{equation}
Using the explicit form of solution of this system it is possible to
construct the general solution of the Monge-Amp\`{e}re equation.

\end{abstract}
\end{titlepage}
\section{Introduction}

The original form of Monge-Amp\`{e}re equation
 in $n$-dimensional space
is the following one \cite{zwi,bed,prog}
\begin{equation}
\det(\phi_{\rho,\sigma})=F(\phi,\phi_\rho,x_\rho)\quad (0\geq \rho,\sigma\geq
n-1)\label{2}
\end{equation}
where $\phi_{\rho,\sigma}\equiv\frac {\partial^2 \phi}{\partial x_\rho
\partial x_\sigma}$ and  $x_\rho$ are the coordinates of the space.

With respect to each of its independent coordinates (\ref{2}) is
an equation of  second order and consequently its general
solution must depend on two arbitrary functions of $n-1$ independent
arguments. We shall show that the homogeneous form of this equation, i.e the
special case where the function $F\equiv 0$, is exactly integrable. In this
case the equation corresponds to the condition that a hypersurface in ${\Bbb R}
^{n+1}$
has zero Gaussian Curvature. In what follows we shall refer to this equation
with $F=0$ as the Monge-Amp\`{e}re equation. There is a deep connection between
this equation and the equation
\begin{equation}
\det\left|\begin{array}{cc}k\phi&\phi_j\\
                           \phi_i&\phi_{ij}\end{array}\right|=0.
\end{equation}

For all $k\neq0$ this equation after a trivial change of unknown function
($\phi^{\frac {k}{ k-1}}$ if $k\neq1,0$, $\log \phi$ if $k=1$) coincides with
the Monge Amp\`ere equation. The particular case $k=0$ corresponds to the
Universal Field Equation \cite{fai} whose general solution was found by means
of a Legendre Transform in \cite{fg}. This solution is connected with the
solution
of the Monge Amp\`ere equation in $n+1$ dimensional space by the substitution
\begin{equation}
\Phi_{MA}(x_1,x_2,\dots x_{n+1})=x_{n+1}\phi_{UFE}(x_1,x_2,\dots x_n)
\end{equation}
Its solution by the methods of the present paper will be given elsewhere.

\section{The general construction}

The vanishing of the  determinant means that
the rows (or  columns) of its $n\times n$ matrix are linearly dependent.
\begin{equation}
\sum_\sigma \phi_{\rho,\sigma} \alpha_\sigma=0 \label{3}
\end{equation}
where $\alpha_\sigma$ are some functions of the coordinates $x_\rho$.
 From this observation it is easy to see that if $\phi$ is homogeneous of
weight one, i.e. $\phi(\lambda x_\rho)=\lambda\phi(x_\rho)$ then by Euler's
theorem on homogeneous functions
\begin{equation}
\sum_\sigma x_\sigma\frac{\partial\phi}{\partial x_\sigma}-\phi =0.
\label{3a}
\end{equation}
Differentiation of this equation with respect to $x_\rho$ leads to an equation
of the form (\ref{3}) and thus implies that  any homogeneous function of weight
one satisfies the  Monge-Amp\`{e}re equation. Thus one class of solutions is
easily obtained.

Let us rewrite (\ref{3}) in equivalent form
\begin{equation}
\frac {\partial {\cal R}}{\partial x_\rho}\equiv\frac {\partial}
{\partial x_\rho}\sum_\sigma \phi_\sigma \alpha_\sigma=\sum_\sigma
\phi_\sigma\frac{\partial \alpha_\sigma}{\partial x_\rho}\label{4}
\end{equation}
and will consider ${\cal R}$ as a function of  $n$ independent variables
$\alpha_\rho$
assuming that $\det (\frac {\partial \alpha_\sigma}{\partial x_\rho})\neq 1)$.

 From (\ref{4}) we obtain immediately
$$
\phi_\sigma=\frac {\partial{\cal R}}{\partial \alpha_\sigma},\quad {\cal
R}=\sum\alpha_\sigma
\frac {\partial {\cal R}}{\partial \alpha_\sigma}
$$
or that ${\cal R}$ is a homogeneous function of variables $\alpha_\rho$
of degree one. So
\begin{equation}
{\cal R}=\alpha_0 R({\alpha_{j}\over \alpha_0}),\quad \phi_0=R-\sum \xi_{j}
\frac {\partial R}{\partial \xi_{j}},\quad \phi_{j}=\frac {\partial R}
{\partial \xi_{j}}\label{5}
\end{equation}
here $\xi_{j}={\alpha_{j}\over \alpha_0}$ and from now on
 all Latin indices take values from 1 up to $n-1$.

The condition of compatibility of (\ref{5}) - the equivalence of
the second mixed derivatives - will give us the dependenced
of the $n-1$ new variables $\xi_{j}$ upon the $n$ previous variables $x_\rho$.

We have firstly
\begin{equation}
\frac {\partial}{\partial x_{k}}\frac {\partial R}{\partial \xi_{j}}=
\frac {\partial}{\partial x_{j}}\frac {\partial R}{\partial \xi_{k}}
\label{6}
\end{equation}
and secondly
\begin{equation}
\frac {\partial}{\partial x_{k}}\frac {\partial \phi}{\partial x_0}=
-\sum\xi_{j}\frac {\partial}{\partial x_{k}}\frac {\partial R}
{\partial \xi_{j}}=-\sum\xi_{j}\frac {\partial}{\partial x_{j}}
\frac {\partial R}{\partial \xi_{k}}= \frac {\partial}{\partial x_0}\frac
{\partial R}{\partial \xi_{k}}\label{7}
\end{equation}
 From the last equality (\ref{7}) we readily obtain
\begin{equation}
\sum_{k}(\frac {\partial \xi_{k}}{\partial x_0}+\sum_{r=1}
^{n-1}\xi_{r} \frac {\partial \xi_{k}}{\partial x_{r}})
\frac {\partial^2 R}{\partial \xi_{k}\partial\xi_{j}}=0\label{8}
\end{equation}
With respect to the variables $q_{k}=\frac {\partial \xi_{k}}
{\partial x_0}+\sum_{r=1}^{n-1}\xi_{r} \frac {\partial
\xi_{k}}{\partial x_{r}}$ (\ref{8}) is a linear system of
algebraic equations which it is possible to solve in the two cases given by the
Fredholm alternative;
\begin{equation}
\det_{n-1}(\frac {\partial^2 R}{\partial \xi_{k}\partial\xi_{j}})=0,\quad
q_{k}\neq 0;\quad \det_{n-1}(\frac {\partial^2 R}{\partial \xi_
{k}\partial\xi_{j}})\neq0,\quad q_{k}=0\label{9}
\end{equation}
We shall consider in this paper the second possibility, hoping
to come back to the first one in futher publications and
restrict ourselves now only to the simplest nontrivial example $n=3$
in  section 5.

\section{Solution of equations of hydrodynamic type}

The system of equations
\begin{equation}
\frac {\partial \xi_{k}}{\partial x_0}+\sum_{k=1}^{n-1}
\xi_{k} \frac {\partial \xi_{k}}{\partial x_{k}}=0 \label{10}
\end{equation}
from the physical point of view is that of the equations of velocity flow
in hydrodynamics in $n-1$ dimensional space.

In the simplest case $n=2$ this is the one component equation
\begin{equation}
\frac {\partial \xi}{\partial x_0}+\xi \frac {\partial \xi}{\partial x_1}=0
\label{11}
\end{equation}
which is connected with the name of Monge  who  first discovered its general
solution in implicit form
$$
x_1-\xi x_0=f(\xi)
$$ where $f(\xi)$ is an arbitrary function of its argument.

The generalisation of this result to the  multidimensional case (\ref{11})
was found by D.B.Fairlie \cite{fai} and consists in the following construction.
Suppose that  the system of $n-1$ equations
\begin{equation}
 x_{j}-\xi_{j} x_0=Q^{j}(\xi_1,\xi_2,\dots\xi_{n-1})\label{12}
\end{equation}
is solved for the unknown functions $\xi_{j}$. Each solution of
this system (\ref{12}) will satisfy (\ref{10}).

This may be proved as follows. After differentiation  of each equation of
(\ref{12}) with respect to the variables $x_{k}$ we obtain
\begin{equation}
\delta_{k,j}=\sum (\frac {\partial Q^{j}}{\partial x_{k}}+
\delta_{j,k})\frac {\partial \xi_{k}}{\partial x_{k}}
\end{equation}
or
\begin{equation}
\frac {\partial \xi_{j}}{\partial x_{k}}=(Ix_0+\frac {\partial Q}
{\partial \xi})^{-1}_{j,k}\label{13}
\end{equation}
The result of differentiation of (\ref{12}) with respect to $x_0$ has
as a corollary
\begin{equation}
-\frac {\partial \xi_{j}}{\partial x_0}=[(Ix_0+\frac {\partial Q}
{\partial \xi})^{-1}\xi)_{j}\label{14}
\end{equation}
The comparision of (\ref{13}) and (\ref{14}) proves that system (\ref{10})
is satisfied.

\section{Continuation.}

Up to now we have used only the condition of compatibility (\ref{7}).
Now let us return to equations (\ref{6}). By simple computations using the
explicit form for derivatives (\ref{13}) it is
easy to show that equations (\ref{6}) may be written in the form
\begin{equation}
\sum\frac {\partial }{\partial \xi_k}
 Q^r\frac {\partial^2 R}{\partial \xi_{r}
\partial\xi_{j}}=\sum\frac {\partial }{\partial \xi_j}Q^j  \frac {\partial^2 R}
{\partial \xi_{r}\partial\xi_{k}}.\label{15}
\end{equation}
or
\begin{equation}
\sum \frac {\partial^2 R}{\partial \xi_{r} \partial \xi_{j}}Q^{j}=
\frac {\partial L}{\partial \xi_{r}} \label{16}
\end{equation}
where $L$ is an arbitrary function of $n-1$ arguments $\xi_{k}$.
With respect to the functions $Q^{j}$ (\ref{16}) is a linear system of
algebraic equations the matrix of which has the determinant not equal
to zero ( the second case of the (\ref{10}).

The solution of (\ref{16}) in terms of Cramers determinants has the usual
form
\begin{equation}
Q^{j}=[(\frac {\partial^2 R}{\partial \xi\partial \xi})^{-1}\frac
{\partial L}{\partial \xi}]^{j}\label{17}
\end{equation}
and so all of the functions $Q^{j}$ are represented in terms of only two
arbitrary functions $R,L$ and their derivatives up to the second and first
orders respectively.

So with the help of equations (\ref{5}) whose integrability is ensured by the
above analysis and (\ref{12}) we obtain the solution
of the Monge-Amp\`{e}re equation which depends upon two arbitrary functions
each of
$n-1$ independent arguments and which fulfils the claim of the introduction
to provide the general solution for the equation under consideration.

\section{The simplest examples}

\subsection{$n=2$}

 From (\ref{5}) we obtain
\begin{equation}
\frac{\partial \phi}{\partial x_0}=R-\xi_1 \frac{\partial R}
{\partial \xi_1},\quad \frac{\partial \phi}{\partial x_1}=\frac{\partial R}
{\partial \xi_1}\label{18}
\end{equation}
The equation (\ref{11}) and its solution are as follows;
\begin{equation}
\frac {\partial \xi_1}{\partial x_0}+\xi_1 \frac {\partial \xi_1}
{\partial x_1}=0,\quad x_1-\xi_1 x_0=F(\xi_1)\label{19}
\end{equation}
In this case the conditions (\ref{5}) are absent and general solution of the
Monge-Amp\`{e}re equation is determined by the pair of arbitrary functions
$R,F$ each of one argument $\xi_1$.

\subsection{$n=3$ - the general case}

\begin{equation}
\frac{\partial \phi}{\partial x_0}=R-\xi_1 \frac{\partial R}{\partial \xi_1}
-\xi_2 \frac{\partial R}{\partial \xi_2},\quad \frac{\partial \phi}
{\partial x_1}=\frac{\partial R}{\partial \xi_1},\quad \frac{\partial \phi}
{\partial x_2}=\frac{\partial R}{\partial \xi_2}\label{20}
\end{equation}
The connection between the  arguments $\xi$ and $x$ keeping in mind (\ref{17})
takes the form
$$
x_1-\xi_1 x_0=Q^1={R_{22}L_1-R_{12}L_2\over R_{22}R_{11}-R_{12}R_{21}}
$$
\begin{equation}
{}\label{21}
\end{equation}
$$
x_2-\xi_2 x_0=Q^2={-R_{12}L_1+R_{22}L_2\over R_{22}R_{11}-R_{12}R_{21}}
$$
where $L_1\equiv\frac{\partial L}{\partial \xi_1}$ and so on.

\subsection{$n=3$-the degenerate case}

The first possibility in (\ref{10}) means $ Det_2(\frac {\partial^2 R}
{\partial \xi}{\partial \xi})=0 $. This exactly the Monge-Amp\`{e}re equation
of first subsection with the solution
$$
\frac{\partial R}{\partial \xi_2}=P-q \frac{\partial P}
{\partial q},\quad \frac{\partial R}{\partial \xi_1}=\frac{\partial P}
{\partial q},\quad \xi_1-q \xi_1=F(q)
$$
Let us substitute all these expressions into (\ref{20}). We obtain
$$
\frac{\partial \phi}{\partial x_0}=R-(\xi_1-\xi_2q) \frac{\partial P}
{\partial q}-\xi_2P,\quad \frac{\partial \phi}{\partial x_2}=P-
q \frac{\partial P}{\partial q},\quad \frac{\partial \phi}
{\partial x_1}=\frac{\partial P}{\partial q}
$$
The condition of compatibility ( the equality of a second mixed derivatives)
gives the equations
$$
q_{x_2}=-q q_{x_1},\quad q_{x_0}=-F(q) q_{x_1},
\quad q q_{x_3}=-F(q) q_{x_2}
$$
which are compatible and have the common solution
$$
-x_1+q x_2-F(q) x_0=\Phi(q)
$$
where $\Phi(q)$ is an arbitrary function of its argument.

We see that the degenerate solution of the Monge-Amp\`{e}re equation is
determined
in this case by three arbitrary functions of one argument.

\section{Conclusion remarks}
The general solution of the the Monge-Amp\`{e}re equation constructed here
reminds one of the situation with regard to linear ordinary second order
equations
when one solution is known; there is a well known procedure for the
construction
of a first order differential equation whose solution yields the general
solution of the original equation. Here something of the same nature is
discovered; an arbitrary homogeneous function of weight one satisfies the
Monge Amp\`{e}re equation automatically. We then construct a set of first order
equations which are integrable, though only implicitly. Their solution then
implies the general solution to our original equation. This construction
suggests that equations with the property that an any homogeneous function of
given degree is a solution may be also solved by this method, for example the
Universal field equation introduced in \cite{fgm}\cite{fg2} whose 3 dimensional
version describes developable surfaces and which admits homogeneous solutions
of degree zero. We intend to return to this matter in a subsequent paper.

\end{document}